\documentclass[twocolumn,
prd, showpacs,
floats]{revtex4}
\usepackage{graphicx}
\usepackage{dcolumn}
\usepackage{bm}
\usepackage{amsmath}

\begin{document}

\title{Torsional Topological Invariants}

\author{H. T. Nieh}
\email{nieh@tsinghua.edu.cn}
\affiliation{Institute for Advanced Study, Tsinghua University, Beijing 100084, China}

\begin{abstract}

Making use of the SO(3,1) Lorentz algebra, we derive in this paper two series of Gauss-Bonnet type identities involving torsion, one being of the Pontryagin type and the other of the Euler type. Two of the six identities involve only torsional tensorial entities and establish $\sqrt{-g}\epsilon^{\mu\nu\lambda\rho}(C^{\alpha\beta}_{~~\mu\nu}+C^{\sigma}_{~\mu\nu}C_{\sigma}^{~\alpha\beta})C_{\lambda\alpha}^{~~\eta}C_{\rho\beta\eta}$ and $\sqrt{-g}\epsilon^{\mu\nu\lambda\rho}\epsilon_{\alpha\beta\gamma\delta}(C^{\alpha\beta}_{~~\mu\nu}+C^{\sigma}_{~\mu\nu}C_{\sigma}^{~\alpha\beta})C^{~\gamma\eta}_{\lambda}C_{\rho~\eta}^{~\delta}$ as purely torsional topological invariants.

\end{abstract}


\date{June 27, 2018}
\maketitle

\vspace{3mm}

\section{Introduction}
Topological considerations have played increasingly active roles in various branches of physics, such as in topological condensed matter physics \cite{Kosterlitz, Thouless, Kane, Bernevig}, and in quantum chaos \cite{Tian}. In the realm of Einstein's Riemannian gravitational theory, there are two well-known topological invariants in terms of the curvature tensor in four dimension. They are the Euler invariant and the Pontryagin invariant. In the advent of the Kibble-Sciama theory of gravitation \cite{Kibble, Sciama}, torsion tensor came into play \cite{Hehl}, and there exists so far one known purely torsional topological invariant, namely the Nieh-Yan invariant \cite{Yan, Zanelli, Nieh 2007} discovered in 1982. This torsional invariant has since been recognized as the generating functional \cite{Mercuri, Date} for the canonical transformation into the Ashtekar variables \cite{Ashtekar} that led to the development of loop quantum gravity \cite{Singh}. It also appeared in studies in torsional chiral anomaly \cite{Zanelli, Soo, Kreimer, Chandia, Peeters, Kimura} and related investigations \cite{Calcgni, Taveras, Mercuri 3}. In this paper, we shall derive two more series of topological invariants involving torsion, one being of the Pontryagin type and the other of the Euler type, with each series consisting of three invariants. Two out of these six invariants contain only torsional tensorial entities.

 In the Einstein-Cartan-Kibble-Sciama theory of gravitation \cite{Kibble, Sciama, Hehl}, the vierbein field $e^{a}_{~\mu}$ and the Lorentz-spin connection field $\omega^{ab}_{~~\mu}$ are the independent basic field variables, where the Latin a, b are the anholonomic Lorentz indices and the Greek $\mu$ the holonomic coordinate index. Geometric entities, such as the metric tensor $g_{\mu\nu}$ and the affine connection $\Gamma_{~\mu\nu}^{\lambda}$ are defined in terms of the basic field variables $e^{a}_{~\mu}$ and $\omega^{ab}_{~~\mu}$. The metric tensor $g_{\mu\nu}$ is defined by
 $$g_{\mu\nu}=\eta_{ab}e_{~\mu}^{a}e_{~\nu}^{b}, \eqno(1) $$
 \noindent with
 $$\eta^{ab}=\eta_{ab}=(1,-1,-1,-1), \eqno(2) $$
 \noindent while the affine connection $\Gamma_{~\mu\nu}^{\lambda}$ is to be defined \cite{Kibble} in such a way that ensures metric compatibility. Defining the covariant derivatives, denoted with semi-column $;$ subscripts, with respect to both local Lorentz transformations and general coordinate transformations for generic $\chi_{a}^{~\lambda}$ and $\chi_{~\nu}^{a}$ according to
$$\chi_{a~;\mu}^{~\lambda}\equiv\chi_{a~,\mu}^{~\lambda}-\omega_{~a\mu}^{b}\chi_{b}^{~\lambda}+\Gamma_{~\nu\mu}^{\lambda}\chi_{a}^{~\nu}, \eqno(3) $$
$$\chi_{~\nu;\mu}^{a}\equiv\chi_{~\nu,\mu}^{a}+\omega_{~b\mu}^{a}\chi_{~\nu}^{b}-\Gamma_{~\nu\mu}^{\lambda}\chi_{~\lambda}^{a}, \eqno(4) $$
\noindent the affine connection is chosen \cite{Kibble} to be
$$\Gamma_{~\mu\nu}^{\lambda}=e_{a}^{~\lambda}(e_{~\mu,\nu}^{a}+\omega_{~b\nu}^{a}e_{~\mu}^{b}), \eqno(5) $$
\noindent where $e_{a}^{~\lambda}$ is the inverse of $e^{a}_{~\lambda}$. We note that the Lorentz indices are raised and lowered by $\eta^{ab}=\eta_{ab}$ while the coordinate indices are lowered and raised by $g_{\mu\nu}$ and its inverse $g^{\mu\nu}$. It then follows from the definition of the affine connection (5) that
$$e_{~\mu;\lambda}^{a}=0, \eqno (6) $$
$$e_{a~;\lambda}^{~\mu}=0, \eqno (7) $$
\noindent and, consequently, with the metric tensor $g_{\mu\nu}$ defined by (1),
$$g^{\mu\nu}_{~~;\lambda}=0, \eqno (8) $$
$$g_{\mu\nu;\lambda}=0. \eqno (9) $$
The affine connection $\Gamma_{~\mu\nu}^{\lambda}$ as defined by (5) is in general not symmetric,
$$\Gamma_{~\mu\nu}^{\lambda}\neq\Gamma_{~\nu\mu}^{\lambda},$$
\noindent giving rise to the torsion tensor $C_{~\mu\nu}^{\lambda}$, which is defined by
$$C_{~\mu\nu}^{\lambda}=\Gamma_{~\mu\nu}^{\lambda}-\Gamma_{~\nu\mu}^{\lambda}. \eqno (10) $$

Define
$$\varpi_{\mu}\equiv\frac{1}{4}\sigma_{ab}\omega^{ab}_{~~\mu}, \eqno (11) $$
\noindent where \cite{Bjorken}
$$\sigma_{ab}=\frac{i}{2}[\gamma_{a},\gamma_{b}], \eqno (12) $$
\noindent with $\{\gamma_{a},\gamma_{b}\}=2\eta_{ab}$.
\noindent The set of matrices $\frac{i}{2}\sigma_{ab}$ satisfy the SO(3,1) Lorentz algebra
$$\frac{i}{2}[\sigma_{ab},\sigma_{cd}]=\eta_{ac}\sigma_{bd}-\eta_{ad}\sigma_{bc}+\eta_{bd}\sigma_{ac}-\eta_{bc}\sigma_{ad}. \eqno (13) $$
The Lorentz curvature $R^{ab}_{~~\mu\nu}$ is defined through
$$\frac{1}{4}\sigma_{ab}R^{ab}_{~~\mu\nu}\equiv\varpi_{\mu,\nu}-\varpi_{\nu,\mu}+i[\varpi_{\mu},\varpi_{\nu}], \eqno (14) $$
\noindent and, as a result of the Lorentz algebra (13), is given by
$$R^{ab}_{~~\mu\nu}=\omega^{ab}_{~~\mu,\nu}-\omega^{ab}_{~~\nu,\mu}-\omega^{ac}_{~~\mu}\omega_{c~\nu}^{~b}+\omega^{ac}_{~~\nu}\omega_{c~\mu}^{~b},\eqno (15)$$
\noindent which has the property
$$
\begin{array}{c}
  R^{\lambda\rho}_{~~\mu\nu}=e_{a}^{~\lambda}e_{b}^{~\rho}R^{ab}_{~~\mu\nu} \\
  =g^{\rho\sigma}(\Gamma^{\lambda}_{~\sigma\mu,\nu}-\Gamma^{\lambda}_{~\sigma\nu,\mu}-\Gamma^{\lambda}_{~\alpha\mu}\Gamma^{\alpha}_{~\sigma\nu}+\Gamma^{\lambda}_{~\alpha\nu}\Gamma^{\alpha}_{~\sigma\mu})
\end{array}
\eqno (16)$$
\noindent where $\Gamma^{\lambda}_{~\mu\nu}$ is defined by (3).

\section{Gauss-Bonnet Type Identities}

Gauss-Bonnet type identities in Riemann-Cartan curved space-time
$$\sqrt{-g}\epsilon^{\mu\nu\lambda\rho}R^{\alpha\beta}_{~~\mu\nu}R_{\alpha\beta\lambda\rho}=\mbox{total derivative}, \eqno (17) $$
$$\sqrt{-g}\epsilon^{\mu\nu\lambda\rho}\epsilon_{\alpha\beta\gamma\delta}R^{\alpha\beta}_{~~\mu\nu}R^{\gamma\delta}_{~~\lambda\rho}=\mbox{total derivative}\eqno (18) $$
\noindent can be simply derived \cite{Nieh 1980} on the basis of the SO(3,1) Lorentz algebra and properties of the Dirac matrices. These identities establish
$$\sqrt{-g}\epsilon^{\mu\nu\lambda\rho}R^{\alpha\beta}_{~~\mu\nu}R_{\alpha\beta\lambda\rho}, \eqno (19) $$
$$\sqrt{-g}\epsilon^{\mu\nu\lambda\rho}\epsilon_{\alpha\beta\gamma\delta}R^{\alpha\beta}_{~~\mu\nu}R^{\gamma\delta}_{~~\lambda\rho} \eqno (20) $$
\noindent as topological invariants. They are, respectively, the Pontryagin and Euler topological invariants.

Based on the SO(4,1) de Sitter algebra, another Gauss-Bonnet type identity
$$\sqrt{-g}\epsilon^{\mu\nu\lambda\rho}R^{AB}_{~~~\mu\nu}R_{AB\lambda\rho}=\mbox{total derivative}, \eqno (21) $$
\noindent where indices A and B take on five valuess (0,1,2,3,5), can be derived \cite{Yan}, establishing
$$\sqrt{-g}\epsilon^{\mu\nu\lambda\rho}R^{AB}_{~~~\mu\nu}R_{AB\lambda\rho} \eqno (22) $$
\noindent as a Pontryagin type topological invariant for the de Sitter group. It is the difference of the SO(4,1) and SO(3,1) Pontryagin invariants, namely (21) and (17), that led to the identity \cite{Yan}
$$\sqrt{-g}\epsilon^{\mu\nu\lambda\rho}(R_{\mu\nu\lambda\rho}+\frac{1}{2}C^{\alpha}_{~\mu\nu}C_{\alpha\lambda\rho})\\
=\partial_{\mu}(-\sqrt{-g}\epsilon^{\mu\nu\lambda\rho}C_{\nu\lambda\rho}), \eqno (23) $$
\noindent establishing
$$\sqrt{-g}\epsilon^{\mu\nu\lambda\rho}(R_{\mu\nu\lambda\rho}+\frac{1}{2}C^{\alpha}_{~\mu\nu}C_{\alpha\lambda\rho}) \eqno (24) $$
\noindent as a torsional topological invariant \cite{Zanelli, Nieh 2007}.

We now use the same method, based on the SO(3,1) algebra and properties of the Dirac matrices, to derive two more series of topological invariants involving torsion, one being of the Euler type and the other of the Pontryagin type. Define $\omega'^{ab}_{~~\mu}$ by
$$\omega'^{ab}_{~~~\mu}\equiv\omega^{ab}_{~~\mu}+\xi C_{\mu}^{~ab}, \eqno (25) $$
\noindent where $\xi$ is an arbitrary parameter, and
$$C_{\mu}^{~ab}=e^{a}_{\lambda}e^{b}_{\rho}C_{\mu}^{~\lambda\rho}, $$
\noindent which is antisymmetric in a and b. It is convenient to introduce the group algebraic notations
$$\begin{array}{rl}\varpi_{\mu}=\frac{1}{4}\sigma_{ab}\omega^{ab}_{~~\mu}\\
C_{\mu}=\frac{1}{4}\sigma_{ab}C^{~ab}_{\mu}\\
\varpi'_{\mu}\equiv\varpi_{\mu}+\xi C_{\mu}=\frac{1}{4}\sigma_{ab}\omega'^{~ab}_{~~~~\mu}\end{array}. \eqno (26) $$
\noindent The curvature tensor $R'^{ab}_{~~\mu\nu}$ corresponding to the connection $\omega'^{ab}_{~~~\mu}$ is defined by
$$\frac{1}{4}\sigma_{ab}R'^{ab}_{~~~\mu\nu}=\bar{R'}_{\mu\nu}, \eqno (27) $$
\noindent where
$$\bar{R'}_{\mu\nu}\equiv\varpi'_{~\mu,\nu}-\varpi'_{~\nu,\mu}+i[\varpi'_{~\mu},\varpi'_{~\nu}], \eqno (28) $$
\noindent On account of the SO(3,1) Lorentz algebra, $R'^{ab}_{~~\mu\nu}$ is explicitly given by
$$R'^{ab}_{~~~\mu\nu}=\omega'^{ab}_{~~~\mu,\nu}-\omega'^{ab}_{~~~\nu,\mu}-\omega'^{ac}_{~~~\mu}\omega'^{~b}_{c~\nu}+\omega'^{ac}_{~~~\nu}\omega'^{~b}_{c~\mu}, \eqno (29) $$
\noindent which can be expressed in terms of the curvature tensor $R^{ab}_{~~\mu\nu}$ and the torsion tensor $C_{\mu}^{~ab}$ as

$$\begin{array}{c}
R'^{ab}_{~~~\mu\nu}=R^{ab}_{~~\mu\nu}+\xi(C^{ab}_{~~\mu\nu}+C_{\lambda}^{~ab}C^{\lambda}_{~\mu\nu})\\
\quad\quad\quad\quad+\xi^{2}(-C_{\mu}^{~ac}C_{\nu c}^{~~b}+C_{\nu}^{~ac}C_{\mu c}^{~~b})\end{array},\eqno (30) $$
\noindent where $C^{ab}_{~~\mu\nu}$ denotes
$$C^{ab}_{~~\mu\nu}=C^{~ab}_{\mu~~;\nu}-C^{~ab}_{\nu~~;\mu},\eqno (31) $$
\noindent with $;$ representing covariant derivative in accordance with the definitions (4) and (5), e.g.,
$$C^{~ab}_{\mu~~;\nu}=C^{~ab}_{\mu~~,\nu}-\Gamma^{\lambda}_{~\mu\nu}C^{~ab}_{\lambda}+\omega^{a}_{~c\nu}C_{\mu}^{~cb}+\omega^{b}_{~c\nu}C_{\mu}^{~ac}. \eqno (32) $$

\section{Ponryagin Type Identities}

We can express the Pontryagin type invariant for $R'^{\alpha\beta}_{~~~\lambda\rho}$ in the following form

$$\begin{array}{c}\sqrt{-g}\epsilon^{\mu\nu\lambda\rho}R'^{\alpha\beta}_{~~~\mu\nu}R'_{\alpha\beta\lambda\rho}\\
=\sqrt{-g}\epsilon^{\mu\nu\lambda\rho}R'^{ab}_{~~~\mu\nu}R'_{ab\lambda\rho}\\
\quad=2\sqrt{-g}\epsilon^{\mu\nu\lambda\rho}{\rm Tr}[\bar{R'}_{\mu\nu}\bar{R'}_{\lambda\rho}]\end{array}, \eqno (33) $$
\noindent where use has been made of
$${\rm Tr}[\sigma_{ab}\sigma_{cd}]=4(\eta_{ac}\eta_{bd}-\eta_{ad}\eta_{bc}). \eqno (34) $$
\noindent With $\bar{R'}_{\mu\nu}$ given by (28), we can express (33) in the form \cite{Nieh 1980}
$$\begin{array}{rl}\sqrt{-g}\epsilon^{\mu\nu\lambda\rho}R'^{\alpha\beta}_{~~~\mu\nu}R'_{\alpha\beta\lambda\rho}\quad\quad\quad\quad\\
=\partial_{\mu}\{8\sqrt{-g}\epsilon^{\mu\nu\lambda\rho}{\rm Tr}[\varpi'_{\nu}\partial_{\lambda}\varpi'_{\rho}+\frac{2i}{3}\varpi'_{\nu}\varpi'_{\lambda}\varpi'_{\rho}]\}\end{array}, \eqno (35) $$

\noindent where use has been made of the fact that $\sqrt{-g}\epsilon^{\mu\nu\lambda\rho}$ is a constant.
With $R'^{ab}_{~~~\mu\nu}$ given by (30), we can expand the left-hand side of (35) as a power series of the parameter $\xi$,
$$\begin{array}{rl}
\sqrt{-g}\epsilon^{\mu\nu\lambda\rho}R'^{\alpha\beta}_{~~~\mu\nu}R'_{\alpha\beta\lambda\rho}\quad\quad\quad\quad\quad\quad\quad\quad
\\
=\sqrt{-g}\epsilon^{\mu\nu\lambda\rho}\{R^{\alpha\beta}_{~~\mu\nu}R_{\alpha\beta\lambda\rho}+\xi 2R^{\alpha\beta}_{~~\mu\nu}(C_{\alpha\beta\lambda\rho}+C_{\sigma\alpha\beta}C^{\sigma}_{~\lambda\varrho})\quad\quad\quad\\
+\xi^{2}[4R^{\alpha\beta}_{~~\mu\nu}C^{~~\sigma}_{\lambda\alpha}C_{\rho\beta\sigma}\quad\quad\quad\quad\quad\quad\quad\\
+(C^{\alpha\beta}_{~~\mu\nu}+C_{\sigma}^{~\alpha\beta}C^{\sigma}_{~\mu\nu})(C_{\alpha\beta\lambda\rho}+C_{\eta\alpha\beta}C^{\eta}_{~\lambda\rho})]\quad\quad\quad\quad\quad\quad\\
+\xi^{3}4(C^{\alpha\beta}_{~~\mu\nu}+C^{~\alpha\beta}_{\sigma}C^{\sigma}_{~\mu\nu})C_{\lambda\alpha}^{~~\gamma}C_{\rho\beta\gamma}\}\quad\quad\quad\quad\quad\quad\end{array}. \eqno (36) $$
\noindent As a power series in $\xi$, and with $\varpi'_{\mu}$ given by (26), the right-hand side of (35) is given by
$$\begin{array}{c}\partial_{\mu}\{8\sqrt{-g}\epsilon^{\mu\nu\lambda\rho}{\rm Tr}[(\varpi_{\nu}\partial_{\lambda}\varpi_{\rho}+\frac{2i}{3}\varpi_{\nu}\varpi_{\lambda}\varpi_{\rho})\\
+\xi (\varpi_{\nu}\partial_{\lambda}C_{\rho}+C_{\nu}\partial_{\lambda}\varpi_{\rho}+2iC_{\nu}\varpi_{\lambda}\varpi_{\rho})\\
+\xi^{2}(C_{\nu}\partial_{\lambda}C_{\rho}+2iC_{\nu}C_{\lambda}\varpi_{\rho})+\xi^{3}C_{\nu}C_{\lambda}C_{\rho}]\}\end{array}. \eqno (37)$$
\noindent We recall that $\varpi_{\mu}$ and $C_{\mu}$ are given in (26). Since the parameter $\xi$ is arbitrary, we equate terms in (36) with corresponding terms of equal power in $\xi$ in (37) and obtain a set of four identities. The identity corresponding to the zeroth power in $\xi$ is the original Gauss-Bonnet identity (16) for the Pontryagin invariant. The other three identities are
$$\begin{array}{rl}\sqrt{-g}\epsilon^{\mu\nu\lambda\rho}R^{\alpha\beta}_{~~\mu\nu}(C_{\alpha\beta\lambda\rho}+C_{\sigma\alpha\beta}C^{\sigma}_{~\lambda\rho})\quad\quad\quad\\
=\partial_{\mu}[4\sqrt{-g}\epsilon^{\mu\nu\lambda\rho}{\rm Tr}(\varpi_{\nu}\partial_{\lambda}C_{\rho}+C_{\nu}\partial_{\lambda}\varpi_{\rho}
+2iC_{\nu}\varpi_{\lambda}\varpi_{\rho})]\end{array}, \eqno (38) $$
$$\begin{array}{rl}\sqrt{-g}\epsilon^{\mu\nu\lambda\rho}[4R^{\alpha\beta}_{~~\mu\nu}C_{\lambda\alpha}^{~~\sigma}C_{\rho\beta\sigma}\quad\quad\quad\\
+(C^{\alpha\beta}_{~~\mu\nu}+C_{\gamma}^{~\alpha\beta}C^{\gamma}_{~\mu\nu})(C_{\alpha\beta\lambda\rho}+C_{\delta\alpha\beta}C^{\delta}_{~\lambda\rho})]\\
=\partial_{\mu}[8\sqrt{-g}\epsilon^{\mu\nu\lambda\rho}{\rm Tr}(C_{\nu}\partial_{\lambda}C_{\rho}+2iC_{\nu}C_{\lambda}\varpi_{\rho})]\end{array},  \eqno (39) $$
$$\begin{array}{rl}\sqrt{-g}\epsilon^{\mu\nu\lambda\rho}(C^{\alpha\beta}_{~~\mu\nu}+C_{\sigma}^{~\alpha\beta}C^{\sigma}_{~\mu\nu})C_{\lambda\alpha}^{~~\gamma}C_{\rho\beta\gamma}\\
=\partial_{\mu}[\frac{4i}{3}\sqrt{-g}\epsilon^{\mu\nu\lambda\rho}{\rm Tr}(C_{\nu}C_{\lambda}C_{\rho})]\end{array}.  \eqno (40) $$

\section{Euler Type Identities}

Let us denote by $\eta_{abcd}$ the totally antisymmetric Minkowski tensor, with
$$\eta_{0123}=-1. \eqno (41) $$
\noindent Because of the relation
$${\rm Tr}[\gamma_{5}\sigma_{ab}\sigma_{cd}]=-4i\eta_{abcd}, \eqno (42) $$
\noindent where
$$\gamma_{5}=i\gamma^{0}\gamma^{1}\gamma^{2}\gamma^{3}, \eqno (43) $$
\noindent we can express the Euler type invariant in the form
$$\begin{array}{rl}\sqrt{-g}\epsilon^{\mu\nu\lambda\rho}\epsilon_{\alpha\beta\gamma\delta}R'^{\alpha\beta}_{~~~\mu\nu}R'^{\gamma\delta}_{~~~\lambda\rho}\\
=\sqrt{-g}\epsilon^{\mu\nu\lambda\rho}\eta_{abcd}R'^{ab}_{~~~\mu\nu}R'^{cd}_{~~~\lambda\rho}\\
=4i\sqrt{-g}\epsilon^{\mu\nu\lambda\rho}{\rm Tr}[\gamma_{5}\bar{R'}_{\mu\nu}\bar{R'}_{\lambda\rho}]\end{array}. \eqno (44) $$
\noindent where $\bar{R'}_{\mu\nu}$ is defined by (28). Substituting (28) into (44), we obtain \cite{Nieh 1980}
$$\begin{array}{rl}\sqrt{-g}\epsilon^{\mu\nu\lambda\rho}\epsilon_{\alpha\beta\gamma\delta}R'^{\alpha\beta}_{~~~\mu\nu}R'^{\gamma\delta}_{~~~\lambda\rho}\quad\quad\quad\\
=\partial_{\mu}\{16i\sqrt{-g}\epsilon^{\mu\nu\lambda\rho}{\rm Tr}[\gamma_{5}(\varpi'_{\nu}\partial_{\lambda}\varpi'_{\rho}+\frac{2i}{3}\varpi'_{\nu}\varpi'_{\lambda}\varpi'_{\rho})]\}\end{array}, \eqno (45) $$
\noindent where use has been made of the relation,
$$[\gamma_{5},\varpi'_{\mu}]=0. \eqno (46) $$
\noindent We expand both sides of (45) as power series of $\xi$. The left-hand side is
$$\begin{array}{rl}\sqrt{-g}\epsilon^{\mu\nu\lambda\rho}\epsilon_{\alpha\beta\gamma\delta}R'^{\alpha\beta}_{~~~\mu\nu}R'^{\gamma\delta}_{~~~\lambda\rho}\quad\quad\quad\\
=\sqrt{-g}\epsilon^{\mu\nu\lambda\rho}\epsilon_{\alpha\beta\gamma\delta}\{R^{\alpha\beta}_{~~\mu\nu}R^{\gamma\delta}_{~~\lambda\rho}\quad\quad\quad\quad\\
+\xi 2R^{\alpha\beta}_{~~\mu\nu}(C^{\gamma\delta}_{~~\lambda\rho}+C_{\sigma}^{~\gamma\delta}C^{\sigma}_{~\lambda\rho})\quad\quad\quad\quad\\
+\xi^{2}[4R^{\alpha\beta}_{~~\mu\nu}C_{\lambda}^{~\gamma\sigma}C_{\rho~\sigma}^{~\delta}+(C^{\alpha\beta}_{~~\mu\nu}\quad\quad\quad\quad\\
+C_{\sigma}^{~\alpha\beta}C^{\sigma}_{~\mu\nu})(C^{\gamma\delta}_{~~\lambda\rho}+C_{\eta}^{~\gamma\delta}C^{\eta}_{~\lambda\rho})]\quad\quad\quad\quad\\
+\xi^{3}4(C^{\alpha\beta}_{~~\mu\nu}+C_{\sigma}^{~\alpha\beta}C^{\sigma}_{~\mu\nu})C^{~\gamma\eta}_{\lambda}C_{\rho~\eta}^{~\delta}\}\quad\quad\quad\quad
\end{array}.  \eqno (47) $$
\noindent The right-hand side of (45), as a power series of $\xi$, is given by
$$\begin{array}{rl}\partial_{\mu}\{16i\sqrt{-g}\epsilon^{\mu\nu\lambda\rho}{\rm Tr}[\gamma_{5}(\varpi_{\nu}\partial_{\lambda}\varpi_{\varrho}+\frac{2i}{3}\varpi_{\nu}\varpi_{\lambda}\varpi_{\varrho})\\
+\xi\gamma_{5}(C_{\nu}\partial_{\lambda}\varpi_{\rho}+\varpi_{\nu}\partial_{\lambda}C_{\rho}+2iC_{\nu}C_{\lambda}C_{\rho})\\
+\xi^{2}\gamma_{5}(C_{\nu}\partial_{\lambda}C_{\rho}+2iC_{\nu}C_{\lambda}\varpi_{\rho})+\xi^{3}\gamma_{5}\frac{2i}{3}C_{\nu}C_{\lambda}C_{\rho}]\}\end{array}.  \eqno (48) $$
\noindent Equating terms in (47) with corresponding terms of the same power in $\xi$ in (48), we obtain four identities. The identity corresponding to the zeroth power in $\xi$ is the original Gauss-Bonnet identity for the Euler invariant (18). The other three identities are the following:

$$\begin{array}{rl}\sqrt{-g}\epsilon^{\mu\nu\lambda\rho}\epsilon_{\alpha\beta\gamma\delta}R^{\alpha\beta}_{~~\mu\nu}(C^{\gamma\delta}_{~~\lambda\rho}+C_{\sigma}^{~\gamma\delta}C^{\sigma}_{~\lambda\rho})\\
=\partial_{\mu}\{8i\sqrt{-g}\epsilon^{\mu\nu\lambda\rho}{\rm Tr}[\gamma_{5}(C_{\nu}\partial_{\lambda}\varpi_{\rho}+\varpi_{\nu}\partial_{\lambda}C_{\rho}\\
+2iC_{\nu}\varpi_{\lambda}\varpi_{\rho})]\}\quad\quad\quad\quad\quad\end{array}, \eqno (49) $$

$$\begin{array}{rl}\sqrt{-g}\epsilon^{\mu\nu\lambda\rho}\epsilon_{\alpha\beta\gamma\delta}[4R^{\alpha\beta}_{~~\mu\nu}C_{\lambda}^{~\gamma\sigma}C_{\rho~\sigma}^{~\delta}\\
+(C^{\alpha\beta}_{~~\mu\nu}+C_{\sigma}^{~\alpha\beta}C^{\sigma}_{~\mu\nu})(C^{\gamma\delta}_{~~\lambda\rho}+C_{\eta}^{~\gamma\delta}C^{\eta}_{~\lambda\rho})]\\
=\partial_{\mu}\{16i\sqrt{-g}\epsilon^{\mu\nu\lambda\rho}{\rm Tr}[\gamma_{5}(C_{\nu}\partial_{\lambda}C_{\rho}+2iC_{\nu}C_{\lambda}\varpi_{\rho})]\}\end{array},  \eqno (50) $$

$$\begin{array}{rl}\sqrt{-g}\epsilon^{\mu\nu\lambda\rho}\epsilon_{\alpha\beta\gamma\delta}(C^{\alpha\beta}_{~~\mu\nu}
+C_{\sigma}^{~\alpha\beta}C^{\sigma}_{~\mu\nu})C^{~\gamma\eta}_{\lambda}C_{\rho~\eta}^{~\delta}\\
=\partial_{\mu}[-\frac{8}{3}\sqrt{-g}\epsilon^{\mu\nu\lambda\rho}{\rm Tr}(\gamma_{5}C_{\nu}C_{\lambda}C_{\rho})]\end{array}. \eqno (51) $$

\section{Purely Torsional Topological Invariants}

In addition to the original Gauss-Bonnet identities (17) and (18) for the Pontryagin and Euler topological invariants, respectively, we have obtained in this paper six additional identities, which do not exist when torsion vanishes. Three of them, namely (38), (39) and (40), are of the Pontryagin type, and the other three, namely (49), (50) and (51), are of the Euler type. Of the six, two of the identities, namely (40) and (51), are special in that they contain only torsion tensorial entities, just like the previously known torsional identity (23). The right-hand side of the two identities (40) and (51) can be easily evaluated. We have the following results for the traces:

$$\epsilon^{\mu\nu\lambda\rho}{\rm Tr}(C_{\nu}C_{\lambda}C_{\rho})=-\frac{i}{2}\epsilon^{\mu\nu\lambda\rho}C_{\nu a}^{~~b}C_{\lambda bc}C_{\rho}^{~ca}, \eqno (52) $$

$$\epsilon^{\mu\nu\lambda\rho}{\rm Tr}(\gamma_{5}C_{\nu}C_{\lambda}C_{\rho})=\frac{1}{4}\epsilon^{\mu\nu\lambda\rho}C_{\nu a}^{~~b}C_{\lambda bc} {^{*}C}_{\rho}^{~ca}, \eqno (53) $$

\noindent where

$$^{*}C_{\rho}^{~ca}\equiv\eta^{cabd}C_{\rho bd}. \eqno (54) $$

\noindent The identities (40) and (51) then become, respectively,

$$\begin{array}{rl}\sqrt{-g}\epsilon^{\mu\nu\lambda\rho}(C^{\alpha\beta}_{~~\mu\nu}+C_{\sigma}^{~\alpha\beta}C^{\sigma}_{~\mu\nu})C_{\lambda\alpha}^{~~\eta}C_{\rho\beta\eta}\\
=\partial_{\mu}[\frac{2}{3}\sqrt{-g}\epsilon^{\mu\nu\lambda\rho}C_{\nu\alpha}^{~~\beta}C_{\lambda\beta\gamma}C_{\rho}^{~\gamma\alpha}]\end{array},  \eqno (55) $$

\noindent and
$$\begin{array}{rl}\sqrt{-g}\epsilon^{\mu\nu\lambda\rho}\epsilon_{\alpha\beta\gamma\delta}(C^{\alpha\beta}_{~~\mu\nu}+C_{\sigma}^{~\alpha\beta}C^{\sigma}_{~\mu\nu})C^{~\gamma\eta}_{\lambda}C_{\rho~\eta}^{~\delta}\\
=\partial_{\mu}[-\frac{2}{3}\sqrt{-g}\epsilon^{\mu\nu\lambda\rho}C_{\nu\alpha}^{~~\beta}C_{\lambda\beta\gamma}{^{*}C}_{\rho}^{~\gamma\alpha}]\end{array}. \eqno (56) $$

\noindent These two identities together with the previously known identity \cite{Yan}

$$\sqrt{-g}\epsilon^{\mu\nu\lambda\rho}(R_{\mu\nu\lambda\rho}+\frac{1}{2}C^{\alpha}_{~\mu\nu}C_{\alpha\lambda\rho})\\
=\partial_{\mu}(-\sqrt{-g}\epsilon^{\mu\nu\lambda\rho}C_{\nu\lambda\rho}) \eqno  $$

\noindent are the three identities containing only torsional tensorial entities, establishing

$$\sqrt{-g}\epsilon^{\mu\nu\lambda\rho}(R_{\mu\nu\lambda\rho}+\frac{1}{2}C^{\alpha}_{~\mu\nu}C_{\alpha\lambda\rho}) \eqno (57) $$

$$\sqrt{-g}\epsilon^{\mu\nu\lambda\rho}(C^{\alpha\beta}_{~~\mu\nu}+C^{\sigma}_{~\mu\nu}C_{\sigma}^{~\alpha\beta})C_{\lambda\alpha}^{~~\eta}C_{\rho\beta\eta} \eqno (58) $$ $$\sqrt{-g}\epsilon^{\mu\nu\lambda\rho}\epsilon_{\alpha\beta\gamma\delta}(C^{\alpha\beta}_{~~\mu\nu}+C^{\sigma}_{~\mu\nu}C_{\sigma}^{~\alpha\beta})C^{~\gamma\eta}_{\lambda}C_{\rho~\eta}^{~\delta} \eqno (59) $$

\noindent as the three purely torsional topological invariants. We remind ourself that $C^{\alpha\beta}_{~~\mu\nu}$ is defined in (31).

We remark that we have applied the same consideration to the case of the SO(4,1) de Sitter algebra \cite{Yan} by combining $\omega'^{a5}_{~~~\mu}=\frac{1}{l}e^{a}_{~\mu}$ with the $SO(3,1)$ connection $\omega'^{ab}_{~~~\mu}$, defined by (25), to form the $SO(4,1)$ de Sitter connection $\omega'^{AB}_{~~~~\mu}$. No additional identity beyond those already obtained is found. These identities may be of use in future investigations involving torsion, such as in finding torsion contributions to the chiral and conformal anomalies, or in the construction of topological field theories. We also remark that the identities (55) and (56) are equally valid for any third rank tensor having the same symmetry property as the torsion tensor, namely, the property of being antisymmetric in two of the tensor indices. In the Kibble-Sciama theory of gravitation, torsion tensor is the only basic entity possessing this property.

\vspace{2mm}
\begin{acknowledgments}

The author would like to thank J. Zanelli for his thoughtful comments.

\end{acknowledgments}

\end{document}